\documentclass{iau}
\usepackage{graphicx}

\usepackage{amsmath}
\usepackage{amssymb}
\usepackage{upgreek}
\usepackage{longtable}
\usepackage[mathscr]{eucal}
\usepackage{graphicx}

\usepackage{color}

\title[Stellar Magnetic Dynamos]
{Stellar Magnetic Dynamos\\ and Activity Cycles}

\author[Nicholas J. Wright]
{Nicholas J. Wright
\affiliation{Centre for Astrophysics Research, University of Hertfordshire, Hatfield, AL10 9AB\\}}

\pubyear{2013}
\volume{302}
\pagerange{001--004}
\jname{Magnetic Fields Throughout Stellar Evolution}
\editors{Editors}
\begin{document}

\maketitle

\begin{abstract}

Using a new uniform sample of 824 solar and late-type stars with measured X-ray luminosities and rotation periods we have studied the relationship between rotation and stellar activity that is believed to be a probe of the underlying stellar dynamo. Using an unbiased subset of the sample we calculate the power law slope of the unsaturated regime of the activity -- rotation relationship as $L_X/L_{bol}\propto Ro^\beta$, where $\beta=-2.70\pm0.13$. This is inconsistent with the canonical $\beta = -2$ slope to a confidence of 5$\sigma$ and argues for an interface-type dynamo. 
We map out three regimes of coronal emission as a function of stellar mass and age, using the empirical saturation threshold and theoretical super-saturation thresholds. We find that the empirical saturation timescale is well correlated with the time at which stars transition from the rapidly rotating convective sequence to the slowly rotating interface sequence in stellar spin-down models. This may be hinting at fundamental changes in the underlying stellar dynamo or internal structure. 
We also present the first discovery of an X-ray unsaturated, fully convective M star, which may be hinting at an underlying rotation - activity relationship in fully convective stars hitherto not observed. 
Finally we present early results from a blind search for stellar X-ray cycles that can place valuable constraints on the underlying ubiquity of solar-like activity cycles.

\end{abstract}

\firstsection
\section{Introduction}

The stellar magnetic dynamo is thought to be driven by the interplay between convection and differential rotation (Parker 1955) in stars with radiative cores and convective envelopes. The observational manifestation of the dynamo is the relationship between rotation and X-ray activity observed in main sequence F, G, K and M stars. This was first quantified by Pallavicini et al. (1981), who found that X-ray luminosity scaled as $L_X \propto (v \, \mathrm{sin} \, i)^{1.9}$, providing the first evidence for the dynamo-induced nature of stellar coronal activity. For very fast rotators the relationship was found to break down with X-ray luminosity reaching a saturation level of $L_X / L_{bol} \sim 10^{-3}$ (Micela et al. 1985), independent of spectral type. This saturation level is reached at a rotation period that increases toward later spectral types (Pizzolato et al. 2003), but it is unclear what causes this. Despite much work there is yet to be a satisfactory dynamo theory that can explain both the solar dynamo and that of rapidly rotating stars and the continued lack of a sufficiently large and unbiased sample has no doubt contributed to this.

We have produced a new catalog of stars with stellar rotation periods and X-ray luminosities, as described in Wright et al. (2011). The catalog includes 824 solar- and late-type stars, 445 field stars and 379 stars in nearby open clusters (ages 40--700~Myrs). The sample was homogenised by recalculating all X-ray luminosities and converting them onto the ROSAT $0.1 - 2.4$~keV band. To minimise biases we removed all sources known to be X-ray variable, those that exhibit signs of accretion, or those in close binary systems. The sample is approximately equally distributed across the colour range $V-K_s = 1.5 - 5.0$ (G2 to M4) with $\sim$30 stars per subtype, dropping to $\sim$10 stars per subtype from F7 to M6. Here we highlight some of the results derived from this sample, and work that followed from it, with particular focus on results that probe the underlying stellar dynamo.

\begin{figure*}
\begin{center}
\includegraphics[height=380pt, angle=270]{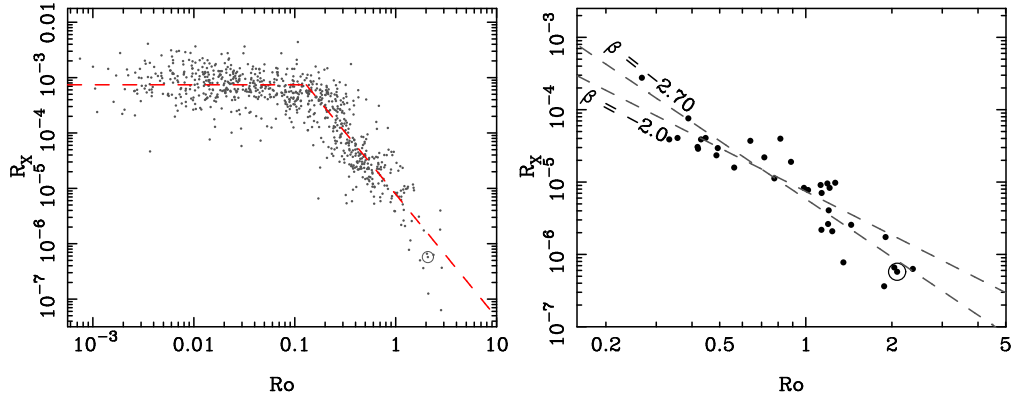}
\caption{$R_X = L_X / L_{bol}$ versus $Ro = P_{rot} / \tau$ for all stars in our sample (left) and for the unbiased sample of 36 stars with unsaturated X-ray emission (right). The Sun is shown with a solar symbol. {\it Left:} The best-fitting saturated and non-saturated activity--rotation relations are shown as a dashed red line. {\it Right:} The log-log ordinary least squares bisector fit, $\beta = -2.70$, and a fit with the canonical slope of $\beta = -2.0$ are shown as dashed lines.}
\label{allstars}
\end{center}
\end{figure*}

\section{Dynamo efficiency in the unsaturated regime}

The relationship between X-ray activity, represented by the ratio of X-ray to bolometric luminosity, $R_X = L_X / L_{bol}$ and the rotation period, parameterised by the spectral-type independent Rossby number, $Ro = P_{rot} / \tau$, the ratio of the rotation period to the convective turnover time (Noyes et al. 1984) is shown in Fig.~\ref{allstars} for all the stars in our sample. The diagram shows the two main regimes of coronal activity: the {\it unsaturated} regime where $R_X$ increases with decreasing $Ro$, and a {\it saturated} regime where the X-ray luminosity ratio is constant with log~$R_X = -3.13 \pm 0.08$. The transition between these two regimes is found to occur at $Ro = 0.13 \pm 0.02$ from a two-part power-law fit (Fig.~\ref{allstars}).

The sample used here suffers from a number of biases, most importantly an X-ray luminosity bias due to the selection only of stars detected in X-rays. To overcome this we used an X-ray unbiased subset of our sample, the 36 Mt. Wilson stars with measured rotation periods (Donahue et al. 1996), all of which are detected in X-rays. While some biases may still exist due to the ability to measure rotation periods, Donahue et al. (1996) conclude that any such biases are unlikely to affect the rotation period distribution. A single-part power-law fit in the linear regime (where $R_X \propto Ro^\beta$) to this sample is shown in Fig.~\ref{allstars} with a fit of $\beta= -2.70 \pm 0.13$. This is a steeper slope than the canonical value of $\beta \simeq -2$ from Pallavicini et al. 1981, though their use of projected rotation velocities instead of rotation periods represents a different relationship than that fitted here. Our slope is inconsistent with the canonical value to a confidence of 5$\sigma$, which argues against a distributed dynamo operating throughout the convection zone, the efficiency of which scales as $Ro^{-2}$ (Noyes et al. 1984), and instead argues for an interface dynamo (e.g., Parker 1993), which has a more complex dependency where $\beta \neq -2$.

\section{The evolution of stellar coronal activity}

\begin{figure*}
\begin{center}
\includegraphics[height=200pt, angle=270]{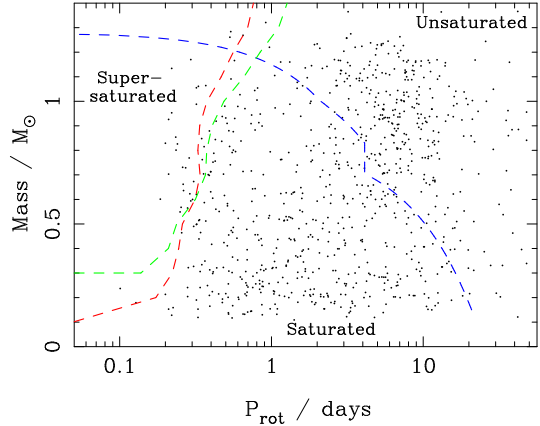}
\caption{The three regimes of coronal X-ray emission shown in Mass-$P_{rot}$ space, the latter of which is an age indicator. All the stars from Wright et al. (2011) are shown as black dots. The empirical saturation threshold of $\mathrm{Ro} < 0.13$ (blue dashed line), and the theoretical supersaturation thresholds for coronal stripping (Jardine \& Unruh 1999, red dashed line) and coronal updrafts (St{\c e}pie{\'n} et al. 2001, green dashed line) are shown.}
\label{regimes}
\end{center}
\end{figure*}

Since all solar and late-type stars are known to emit X-rays (e.g., Vaiana et al. 1981, Wright et al. 2010, Hynes et al. 2012), and X-ray emission is dependent on rotation, which is itself an age indicator (stars spin down as they age and lose angular momentum, Skumanich 1972), stellar coronal activity should evolve in a well-defined manner as stars age. This is shown in Figure~\ref{regimes}, which can be used to follow the coronal evolution of a star from super-saturated, through the saturation regime, to the unsaturated regime. A number of interesting features are revealed by this diagram, notably that F-type stars do not pass through a saturated regime, going straight from super-saturated to unsaturated coronal emission. This explains why previous authors (e.g., Pizzolato et al. 2003) noted a lower $R_X$ saturation level for F-type stars than for other solar- and late-type stars. It now appears that these stars were not saturated but super-saturated, and therefore their X-ray emission levels should not be considered indicative of saturated X-ray emission. It is now the case that {\it all} solar- and late-type stars exhibit the same (within the uncertainties) level of saturated X-ray emission of log~$R_X = -3.13$, suggesting a common mechanism for coronal saturation. Furthermore we note a correlation between the empirical timescale for coronal saturation and the timescale at which stars transition from the rapidly rotating convective sequence to the slowly rotating interface sequence in stellar spin-down models (e.g., Barnes 2003). That these two critical changes in stellar rotation and dynamo activity occur at similar times may be hinting at a common origin from a fundamental change in the underlying stellar dynamo or internal structure.

\section{Slowly rotating fully convective stars}

Late-type M dwarfs become fully convective at approximately a spectral type of M4-5 and as such should not have the interface region between the radiative core and convective envelope (known as the {\it tachocline}) that is believed to be necessary for the generation of an $\alpha\Omega$ dynamo (Parker 1993). As such it is expected that fully convective M dwarfs should not exhibit a rotation -- activity relationship, though this is hard to verify since most late-type stars rotate very rapidly and have large convective turnover times, such that they are all found in the saturated regime of coronal X-ray emission. To fully test this idea we are observing a number of slowly rotating and fully convective M dwarfs with the {\it Chandra} X-ray Observatory to search for evidence of a rotation -- activity relationship. Only one target has been observed so far and was observed to have log~$R_X \sim -4.5$, significantly below the saturation level of log~$R_X \sim -3$. This is the first evidence of non-saturated X-ray emission in a fully convective M-type star and might provide the first evidence for a rotation -- activity relationship in fully convective stars with considerable implications for the stellar dynamo at work in fully convective stars.

\section{A blind search for stellar X-ray cycles}

While chromospheric activity cycles are a well-studied phenomenon amongst solar- and late-type stars, only a handful of coronal activity cycles have been identified and studied (e.g., Favata et al. 2008). To rectify this situation we are performing blind searches for stellar X-ray cycles using deep and long-baseline X-ray observations. The initial dataset used for this project is the {\it XMM-Newton} archive, from which Hoffman et al. (2012) identified nine stars in six fields with data of sufficient depth to extract reliable photometry and long-enough baselines to identify cyclic behaviour. Using a Lomb-Scargle periodogram they searched for cycles, but none were found. From Monte Carlo simulations they simulated their detection capabilities and, assuming a uniform distribution of cycle periods and strengths over the domain searched, conclude with 95\% confidence that $<$72\% of moderately active stars have 5--13 year coronal cycles with an amplitude similar to that of the Sun. Further ongoing studies will provide better constraints on the ubiquity of coronal cycles in solar- and late-type stars.

\section{Summary}

A new catalog of stars with measured X-ray luminosities and rotation periods is used to study the rotation -- activity relation. The power-law slope of the unsaturated regime is fit as $\beta = -2.7 \pm 0.13$, inconsistent with the canonical $\beta = -2$ value with a confidence of 5$\sigma$ and arguing for an interface-type dynamo. We present the first discovery of an X-ray unsaturated fully convective M-type star, which may be hinting an an underlying rotation -- activity relationship in fully convective stars, and present early results from a blind search for stellar X-ray cycles.

\end{document}